\documentstyle[sprocl]{article}

\def\s0#1#2{\mbox{\small{$ \frac{#1}{#2} $}}}
\def\0#1#2{\frac{#1}{#2}}
\def\de{\delta}

\def\eq#1{(\ref{#1})}

\def\beq{\begin{equation}}
\def\eeq{\end{equation}}
\def\bea{\begin{eqnarray}}
\def\eea{\end{eqnarray}}

\begin{document}

\begin{flushright}
CERN-TH/2000-310\\[7ex]
\end{flushright}

\title{Aspects of semi-classical transport theory for QCD
  \footnote{Talk presented at Strong and Electroweak Matter
    (SEWM2000), Marseille, France, 14-17 June 2000.  Based on
    ref.~\cite{LM}${}^{\mbox{-}}\,$\cite{LM4} done in collaboration
    with Cristina Manuel.}  }

\author{Daniel F. Litim \footnote{Daniel.Litim@cern.ch}}

\address{Theory Division, CERN, CH-1211 Geneva 23, Switzerland.}

\maketitle
 
\abstracts{ We discuss some aspects of a recently proposed
  semi-classical transport theory for QCD plasmas based on coloured
  point particles.  This includes the derivation of effective
  transport equations for mean fields and fluctuations which relies on
  the Gibbs ensemble average.  Correlators of fluctuations are
  interpreted as collision integrals for the effective Boltzmann
  equation. The approach yields a recipe to integrate-out
  fluctuations. Systematic approximations (first moment, second
  moment, polarisation approximation) based on a small plasma
  parameter are discussed as well.  Finally, the application to a hot
  non-Abelian plasma close to thermal equilibrium is considered and
  the consistency with the fluctuation-dissipation theorem
  established.}

\section{Introduction}

Reliable theoretical tools to describe transport phenomena of hot or
dense QCD plasmas in- or out-of equilibrium are at the basis for an
understanding of the quark-gluon-plasma. This extreme state of matter
where quarks and gluons are no longer confined is expected to be
produced within the up-coming heavy ion experiments at RHIC and LHC.

In the present contribution a recently proposed semi-classical method
to derive effective transport equations for QCD is reviewed, developed
in collaboration with Cristina Manuel in
ref.~\cite{LM}${}^{\mbox{-}}\,$\cite{LM4}.  The interest in a
formalism based on a classical transport theory \cite{Heinz} is that
main properties of a hot {\it quantum} plasma can already be
understood within simple {\it classical} terms.  Indeed, the soft
non-Abelian gauge fields -- those having a huge occupation number --
can be treated as classical {\it fields}, while the hard gauge field
modes and the quarks can be treated as classical {\it particles}.
This approach has been known since long for Abelian plasmas \cite{K},
but a consistent extension to the non-Abelian case was longtime
missing.

Here, we discuss some conceptual aspects of this method, and in
particular the step from a microscopic to a macroscopic (kinetic)
formulation, the close relationship between the ensemble average and
the notion of coarse-graining, and systematic expansions in a small
plasma parameter or gauge coupling. The latter is at the basis for the
integrating-out of fluctuations and the derivation of collision
integrals. For a hot plasma close to equilibrium it has been shown
that the HTL effective theory follows to leading order in an expansion
in moments \cite{KLLM}. We review how B\"odeker's effective theory
\cite{DB} is recovered at second order in this expansion. Finally, the
compatibility with the fluctuation-dissipation theorem is established
\cite{LM3}.

\section{Classical particles and fields}

The starting point for a classical transport theory is to consider an
ensemble of classical point particles interacting through classical
non-Abelian gauge fields. Such an approach is known since long for
Coulomb plasmas of charged point particles interacting through
photons \cite{K}. The new ingredience in the case of QCD is that the particles
have to carry a non-Abelian colour charge $Q^a$, where the colour
index runs from $a=1$ to $N^2-1$ for a SU($N$) gauge group. These
particles interact self-consistently amongst each others, that is,
through the classical gauge fields created by the particles. Their
classical equations of motion have first been given by Wong
\cite{Wong},
\beq
m\0{d{\hat x}^\mu}{d\tau}={\hat p}^\mu \ ,\quad
m\0{d{\hat p}^\mu}{d\tau}=g {\hat Q}^a F_a^{\mu\nu} {\hat p}_\nu \ ,\quad
m\0{d{\hat Q}^a}{d\tau}=-g f^{abc} {\hat p}^\mu A^{b}_{\mu}{\hat Q}^c \
.\label{dQ}
\eeq
Here, $A_\mu$ denotes the gauge field, $F^{a}_{\mu\nu}=\partial_\mu
A^{a}_\mu-\partial_\nu A^{a}_\mu+ gf^{abc}A^{b}_\mu A^{c}_\nu$ the
corresponding field strength and $\hat z\equiv (\hat x,\hat p,\hat
Q)(\tau)$ the world line of a particle. 
The particles interact self-consistently through classical Yang-Mills fields,
\beq\label{YM} 
D_\mu F^{\mu\nu}=J^\nu\ , 
\eeq 
where the current $J$ is the sum of the currents for the individual
particles, $J=g\sum_i\int d\tau \delta(x-\hat x_i)d\hat x_i/d\tau$.

To further a kinetic description it is useful to introduce a
microscopic one-particle distribution function $f(z)\sim\sum_i \int d\tau
\delta[z-\hat z_i(\tau)]$ which describes an ensemble of such
particles. Making use of (\ref{dQ}), its kinetic equation is
\cite{Heinz}
\beq
\label{NA-f}
p^\mu\left(\0{\partial}{\partial x^\mu} - g f^{abc}A^{b}_\mu
Q^c\0{\partial}{\partial Q^a} -gQ_aF^{a}_{\mu\nu}\0{\partial}{\partial
p_\nu}\right) f(x,p,Q)=0 \ .  
\eeq 
This Boltzmann equation is collisionless, since $df/d\tau=0$
(Liouville's theorem). However, it contains {\it effectively}
collisions due to the long range interactions as shall become clear in
the sequel. 
The Wong equations transform covariantly under gauge transformations 
which entails that $f$ and (\ref{NA-f}) are gauge invariant~\cite{KLLM}.
Eq.~(\ref{NA-f}) is completed by the Yang-Mills equation
(\ref{YM}), where the current $J$ becomes now a functional $J[f]$
of the one-particle distribution function. 
These equations constitute the basis for the construction
of a semi-classical kinetic theory.

\section{Kinetic theory}

Within the semi-classical approach all informations about properties
of the QCD plasma are given by the microscopic dynamical equations
(\ref{YM}) and (\ref{NA-f}). However, for most situations not all
microscopic informations are of relevance. Of main physical interest are
the characteristics of the QCD plasma at macroscopic scales. This
includes quantities like damping rates, colour conductivities or
screening lengths within the kinetic regime, or transport coefficients
like shear or bulk viscosities within the hydrodynamic regime. The
microscopic length scales, like typical inter-particle distances, are
much smaller than such macroscopic scales.

Given the statistical ensemble -- representing the state of the system
-- its macroscopic properties follow as functions of the fundamental
parameters and the interactions between the particles. Within kinetic
theory, the basic `macroscopic' quantity is the one-particle
distribution function from which all further macroscopic observables
can be derived. The aim of a kinetic theory is to construct -- with as
little restrictions or assumptions as possible -- a closed set of
transport equations for this distribution function.  Such an approach
assumes implicitly that the `medium', described by the distribution
function, is continuous. If the medium is {\it not} continuous,
stochastic fluctuations due to the particles should be taken into
account as well, and their consistent inclusion leads to a coupled set
of {\it effective} transport equations for correlators of fluctuations
and the one-particle distribution function \cite{K}.  In this
classical picture, $f$ is a deterministic quantity once all initial
conditions have been fixed. Unfortunately, this is not possible, and
$f$ has to be considered as a {\it stochastic} (fluctuating) quantity
instead.  The random fluctuations of the distribution function are at
the root of the dissipative character of the effective transport
theory.

\section{Coarse graining and ensemble average}

The step from a microscopic to a macroscopic (kinetic) description
requires an appropriate definition of macroscopic quantities in the
first place. There are (at least) two possible options. The first one
consists in taking both volume and time averages of the microscopic
distribution function $f$, the non-Abelian fields and the Boltzmann
equation (\ref{NA-f}) over characteristic physical length and time
scales \cite{LM3}. Clearly, the notion of a characteristic physical
scale depends on the particular problem studied. A reasonable choice
for a coarse graining scale $r_{\rm ph}$ 
has to met some few criteria. The coarse graining scale $r_{\rm
  ph}$ should be sufficiently large such that the number of particles
$N_{\rm ph}$ contained in $r^3_{\rm ph}$ is $\gg 1$. This is at the
basis for a plasma description in the first place, because $N_{\rm
  ph}\gg 1$ ensures that many particles interact coherently within a
coarse-graining volume (quasi-particle behaviour).  
The inverse of $N_{\rm ph}$ is also known as
the {\it plasma parameter} $\epsilon$. In addition, $r_{\rm
  ph}$ should be larger than a typical two-particle correlation length
$r_{\rm corr}$. This ensures that two-particle correlators $g_2$ can
be neglected. Finally, $r_{\rm ph}$ should be smaller than
the macroscopic scales of the plasma to be investigated, like typical
relaxation lengths $r_{\rm rel}$. This way it is guaranteed that the
scales of interest are not washed-out. Such coarse grainings remove
(irrelevant) microscopic information and thereby modify the transport
equation which is expected to become of the Boltzmann-Langevin type,
\beq \label{NA-f-cg} 
p^\mu\left(\bar D_\mu-gQ_a\bar F^{a}_{\mu\nu}{\partial_p^\nu}\right) \bar f 
=C[\bar f]+\zeta \ .  
\eeq 
Here, we introduced $D_\mu[A]f \equiv [\partial_\mu-g f^{abc}Q_c
A_{\mu,b}{\partial ^Q_a}]f$ and the shorthands
$\partial_\mu\equiv{\partial }/{\partial x^\mu}$,
$\partial^p_\mu\equiv{\partial }/{\partial p^\mu}$ and $\partial
^Q_a\equiv{\partial }/{\partial Q^a}$.  The new terms on the r.h.s.~are
the collision integral $C[\bar f]$
and a related source for stochastic noise $\zeta$ on the r.h.s.~of
(\ref{NA-f-cg}) are due to the fact that the Boltzmann equation
(\ref{NA-f}) is quadratic and cubic in the fields. Physically
speaking, these terms arise because the coarse-graining integrates-out
the short range modes within a coarse-graining volume. The
interactions of such modes can result into additional effective
interactions for the remaining long range modes.

A second route for obtaining kinetic equations
consists in taking an ensemble average of the
microscopic transport equation. In the present context we consider
particles in a phase space, hence the appropriate average is the Gibbs
ensemble average henceforth denoted by $\langle\cdots\rangle$. All
statistical informations of the system are then contained in the
correlators $\langle\delta f\cdots\delta f\rangle$ with $\delta
f=f-\langle f\rangle\equiv f-\bar f$ 
denoting the statistical fluctuations of $f$
about its mean value. The most important equal-time correlator is the
quadratic one given by
\beq\label{BasicGibbs} 
\langle \delta f_{{\bf x},p,Q} \, \delta f_{{\bf x}',p',Q'} \rangle 
= (2\pi)^3\delta^{(3)}({\bf x}-{\bf x}')
          \delta^{(3)}({\bf p}-{\bf p}') 
          \delta (Q- Q')\, \bar f +{\cal O}(g_2)
\eeq 
where we have neglected higher order corrections due to the
two-particle correlation function $g_2$. Similar formula hold for
higher order equal-time correlation function. Notice that
(\ref{BasicGibbs}) does not require further informations about the
state of the system and applies equally to systems in- or out-of
equilibrium. While (\ref{BasicGibbs}) has been proven rigorously for
particles with classical statistics, its generalisation to the case of
quantum statistics is known only for weakly interacting particles and
consists in replacing 
\beq\label{replace}
\bar f \to \bar f(1\pm \bar f)
\eeq 
in (\ref{BasicGibbs}) for Bose-Einstein $(+)$ or Fermi-Dirac $(-)$ statistics. 

\section{Effective transport equations}

We now apply the Gibbs ensemble average to the Boltzmann equation
(\ref{NA-f}) in order to derive an effective kinetic equation of the
form (\ref{NA-f-cg}). To that end, we split $f$ into its mean value
and a fluctuating part
\beq f(x,p,Q)=\label{NA-delta-f} 
{\bar f}(x,p,Q) + \de f(x,p,Q) \ ,
\eeq
where ${\bar f} = \langle f \rangle$. The mean value of the
statistical fluctuations vanish by definition, $\langle \de f \rangle
=0$. This separation into mean fields and statistical, random
fluctuations corresponds effectively to a split into long vs.~short
wavelength modes associated to the mean fields vs.~fluctuations.  In
addition to (\ref{NA-delta-f}) we split the gauge field into a mean
field part and fluctuations,
\beq 
A^\mu(x)= {\bar A}^\mu(x) + a^\mu(x) \ ,\label{NA-delta-a} 
\eeq 
with ${\bar A} = \langle A \rangle$ and $\langle a \rangle=0$. This
split should been seen on a different footing as the split
(\ref{NA-delta-f}) because the Gibbs ensemble average is defined for
the particle-like degrees of freedom in the first place. One may 
wonder whether (\ref{NA-delta-a}) -- and the subsequent split of the
dynamical equations -- is compatible with gauge symmetry because the
gauge field transforms non-linearly. The situation is very similar to
the background field method used in path integral formulations of QFT 
and the compatibility of (\ref{NA-delta-a}) with gauge symmetry follows 
along the same lines. For more details, see ref.~\cite{LM2}.
 
Coming back to the initial set of dynamical equations (\ref{YM}) and
(\ref{NA-f}), we perform the ensemble average to find 
\beq
p^\mu\left(\bar D_\mu- gQ_a \bar F^a_{\mu\nu} 
 \partial _p^\nu\right) f=\left\langle\eta\right\rangle
+ \left\langle\xi\right\rangle \ . \label{NA-1}
\eeq
for the effective transport equation, and
\beq
\bar D_\mu \bar F^{\mu\nu}  + \left\langle J_{\mbox{\tiny fluc}}^{\nu}
\right\rangle =\bar J^\nu \ .\label{NAJ-1} 
\eeq
for the effective Yang-Mills equations.
In \eq{NA-1} and \eq{NAJ-1}, we collected all terms quadratic or cubic in the
fluctuations into the functions
\begin{eqnarray}
\eta &\equiv & g Q_a\, p^\mu 
\left[ (\bar D_\mu a_\nu- \bar D_\nu a_\mu)^a+ g f^{abc}a^b_\mu
a^c_\nu\right]\, \partial _p^\nu  \delta f(x,p,Q)  \ , \label{NA-eta}\\
\xi &\equiv & gp^\mu f^{abc}Q^c\left[\partial ^Q_a a_\mu^b\ 
\delta f(x,p,Q)\ + g a_\mu^a a_\nu^b\partial^\nu_p\bar f(x,p,Q)\right]
,\label{NA-xi}\\
J_{\mbox{\tiny fluc}}^{a,\nu}&\equiv & g \left[f^{dbc} 
\bar D^{\mu}_{ad}  a_{b,\mu} 
a_c^\nu  + f^{abc}  a_{b,\mu}\, \left( (\bar D^\mu a^\nu-\bar D^\nu
a^\mu)_c+ g f^{cde}a^\mu_d a^\nu_e \right) \right] .\quad \label{NA-Jfluc}
\end{eqnarray}
The set of equations \eq{NA-1} -- \eq{NA-Jfluc} is not yet closed as
they still involve correlators of fluctuations. Hence, we need their
dynamical equations as well. Subtracting
(\ref{NA-f}) from (\ref{NA-1}) we find for the dynamics of $\delta f$
\begin{eqnarray}
p^\mu\left(\bar D_\mu - gQ_a\bar F^a_{\mu\nu}\partial _p^\nu\right)\delta f
&=&
g Q_a(\bar D_\mu a_\nu-\bar D_\nu a_\mu)^a p^\mu \partial ^p_\nu\bar f
\nonumber \\ && 
+ gp^\mu a_{b,\mu} f^{abc} Q_c\partial ^Q_a \bar f 
+\eta  + \xi - \left\langle\eta+\xi\right\rangle\, . \label{NA-2}
\end{eqnarray}
Subtracting
(\ref{YM}) from (\ref{NAJ-1}) yields the corresponding equation for 
the gauge field fluctuations, to wit
\beq
\left(\bar D^2 a^\mu-\bar D^\mu(\bar D_\nu a^\nu)\right)^a
+ 2 gf^{abc}
\bar F_b^{\mu\nu}a_{c,\nu}+J_{{\mbox{\tiny fluc}}}^{a,\mu}-\left\langle
J_{{\mbox{\tiny fluc}}}^{a,\mu}\right\rangle
=\delta J^{a,\mu} \ .\label{NAJ-2}
\eeq
Eqs.~(\ref{NA-2}) and (\ref{NAJ-2}) are the master equations for all
higher order correlation functions of fluctuations, 
e.g.~multiplying (\ref{NA-2})
by $\delta f$ and taking the ensemble average yields the Boltzmann 
equation for the correlator $\langle \delta f\delta f\rangle$.
This hierarchy of
dynamical equations is similar to the BBGKY hierarchy. All the
equal-time correlators of $\delta f$ can be derived from the basic
definition of the Gibbs ensemble average in phase space and serve as
initial conditions \cite{LM2}.

We remark that the set of dynamical equations (\ref{NA-1}) -- (\ref{NAJ-2})
is equivalent to the original set of microscopic equations. It is
exact in that no further approximations -- apart from the assumptions
leading to (\ref{dQ}) and (\ref{YM}) -- have been performed so far.
Furthermore, it can be applied to out-of-equilibrium situations simply
because the Gibbs ensemble average is not bound to a plasma
close-to-equilibrium.

\section{Collision integrals}

Let us comment on the terms $\eta, \xi$ and $J_{\mbox{\tiny fluc}}$ in
the dynamical equations.  In the effective Boltzmann equation, the
functions $\langle\eta\rangle$ and $\langle\xi\rangle$ appear only
after the splitting (\ref{NA-delta-f}) and (\ref{NA-delta-a}) has been
performed. These terms are qualitatively different from those already
present in the transport equation.  The correlators
$\langle\eta\rangle$ and $\langle\xi\rangle$ are interpreted as
effective collision integrals of the macroscopic Boltzmann equation.
The fluctuations in the distribution function of the particles induce
fluctuations in the gauge fields, while the gauge field fluctuations,
in turn, induce fluctuations in the motion of the quasi-particles. In
the present formalism, the correlators of statistical fluctuations
have the same effect as collisions. This yields a precise recipe for
obtaining collision integrals within semi-classical transport theory.

In the Abelian limit the transport equations reduce to the known set
of kinetic equations for Abelian plasmas and only the collision
integral $\langle\eta\rangle$ survives.  Here it is known that
$\langle\eta\rangle$ can be explicitly expressed as the Balescu-Lenard
collision integral \cite{K}. This substantiates the correspondence between
fluctuations and collisions in an Abelian plasma.

At the same time we observe the presence of stochastic noise in the
effective equations. The noise originates in the source fluctuations
of the particle distributions and induces {\it field-independent}
fluctuations to the gauge fields. The corresponding terms in the
effective Boltzmann equations are therefore $\eta$, $\xi$ and
$J_{\mbox{\tiny fluc}}$ at vanishing mean field or mean current.

Finally, we observe the presence of a fluctuation-induced current
$\langle J_{\mbox{\tiny fluc}} \rangle$ in the effective Yang-Mills
equation for the mean fields. This current, due to its very nature,
stems from the induced correlations of gauge field fluctuations. It
vanishes identically in the Abelian case. While the collision
integrals are linear in the quasi-particle fluctuations, the induced
current only contains the gauge field fluctuations. As the
fluctuations of the one-particle distribution function are the basic
source for fluctuations, we expect that a non-vanishing induced
current will appear as a subleading effect.

In order to find explicitly the collision integrals, noise sources or
the fluctuation-induced currents for non-Abelian plasmas, one has to
solve first the dynamical equations for the fluctuations in the
background of the mean fields.  This step amounts to `integrating-out'
fluctuations. In general, this is a difficult task, in particular due
to the non-linear terms present in $\eta, \xi$ and $J_{\mbox{\tiny
    fluc}}$. This will only be possible within some approximations.

\section{Plasma parameter and moment expansions}

When it comes to solving the kinetic equations -- and in particular
the fluctuation dynamics -- it is necessary to identify a small
expansion parameter which allows for systematic approximations of the
hierarchy of dynamical equations for correlation functions.  The
non-Abelian plasma is characterised by two dimensionless parameters,
the gauge coupling $g$ and the plasma parameter $\epsilon$. A small
plasma parameter is mandatory for a kinetic description to be viable.
For a classical plasma, $\epsilon$ is an independent parameter related
to the mean particle number and the gauge coupling. For a quantum
plasma, the mean particle number cannot be fixed arbitrarily, and
$\epsilon$ scales proportional to the gauge coupling as $\epsilon\sim
g^3\ll 1$, explaining why a kinetic description for quantum plasmas is
intimately linked to the weak coupling limit.

Here, two systematic approximation schemes are outlined: an
expansion in moments of the fluctuations and an expansion in a small
gauge coupling.  Although they have distinct origins in the first
place they are intimately linked due to the requirements of gauge
invariance.

A systematic perturbative expansion in powers of $g$ can
be done because the differential operator appearing in the effective
Boltzmann equation (\ref{NA-1}) admits such an expansion.  The force
term $g\,p^\mu Q_a\bar F^a_{\mu\nu}\partial _p^\nu$ is suppressed by a
power of $g$ as compared to the leading order term $p^\mu\bar D_\mu$.
In this spirit, we expand 
\beq\label{f-g} 
\bar f = \bar f^{(0)} +g\, \bar f^{(1)} +\ldots 
\eeq 
This is at the basis for a systematic organisation of the dynamical
equations in powers of $g$.\cite{KLLM}  To leading order, this concerns in
particular the cubic correlators appearing in $\langle\eta\rangle$ and
$\langle J_{\mbox{\tiny fluc}}\rangle$. They are suppressed by a power
of $g$ as compared to the quadratic ones. 
At the same time, the quadratic correlator $\sim
f_{abc}\langle a^b\,a^c\,\rangle$ within $\langle\xi\rangle$ is also
suppressed by an additional power of $g$ and should be suppressed to
leading order.

An expansion in moments of the fluctuations has its origin in the
framework of kinetic equations, which describe the coherent behaviour
of the particles within some physically relevant volume, described by
the plasma parameter $\epsilon$.  The fluctuations in the number of
particles $\sim N^{-1/2}_{\rm ph}$ become arbitrarily small if the
physical volume -- or the number of particles contained in it -- can
be made arbitrarily large.  For realistic situations, both of them are
finite. Still, the fluctuations remain at least parametrically small
and suppressed by the plasma parameter.  Hence, the underlying
expansion parameter for an expansion in moments of fluctuations is a
small {plasma parameter}
\beq 
        \epsilon \ll 1\ .  
\eeq 
The leading
order approximation in an expansion in moments assumes $\epsilon\equiv 0$.
This is the {\it first
moment approximation} which consists in imposing 
\beq\label{1st} 
f=\bar f\ ,\quad {\rm or}\quad \de f\equiv 0\ , 
\eeq 
and neglects fluctuations throughout. Sometimes it is referred to as
the {\it mean field} or {\it Vlasov approximation}. It leads to a closed
system of equations for the mean one-particle distribution function
and the gauge fields. In particular, the corresponding Boltzmann
equation is dissipationless.

Beyond leading order, the {\it second moment approximation} takes into
account the corrections due to correlators up to quadratic order in
the fluctuations $\langle \de f\de f\rangle$. All higher order
correlators like $\langle \de f_1\de f_2\ldots\de f_n\rangle=0$ for
$n>2$ are neglected within the dynamical equations for the mean fields
and the quadratic correlators. This approximation is viable if the
fluctuations remain sufficiently small. We remark that the second
moment approximation no longer yields a closed system of dynamical
equations for the one-particle distribution function and quadratic
correlators because the initial conditions for the evolution of
correlators, which are given by the equal-time correlation functions
as derived from the Gibbs ensemble average, involve the two-particle
correlation functions $g_2$. Hence, we have to require that
two-particle correlators remain sufficiently
small as compared to products of
one-particle distribution functions. This
approximation is known as the {\it approximation of second correlation
  functions}, sometimes also referred to as the {\it polarisation
  approximation} \cite{K}.

For the dynamical equations of the fluctuations these
approximations imply that the terms non-linear in the fluctuations
should be neglected to leading order, setting
\beq\label{polarisation}
\eta-\langle\eta \rangle=0\ ,\quad
\xi -\langle \xi \rangle=0\ ,\quad
J_{{\mbox{\tiny fluc}}}-\langle J_{{\mbox{\tiny fluc}}}\rangle=0\ .
\eeq
It permits truncating the infinite
hierarchy of equations for the mean fields and the correlators of
fluctuations down to a closed system of differential equations for
both mean quantities and quadratic correlators. 
The polarisation approximation is the minimal
choice necessary to genuinely describe dissipative processes, because
it takes into account the feed-back of stochastic fluctuations within
the particle distribution function.
 
Finally we notice that small fluctuations $\de f$ about $\bar f$ in
(\ref{f-g}) are already of the same order of magnitude as the small
corrections from $g\, \bar f^{(1)}$ once $g\bar f^{(1)} /\bar f^{(0)}
\sim N^{-1/2}_{\rm ph}\ll 1$. This means that fluctuations do
contribute to the effective kinetic equations even if only small
departures from the mean value $g\bar f^{(1)}\ll \bar f^{(0)}$ are
considered. This shall become more explicit at the example of a plasma
close to thermal equilibrium.

\section{Thermal plasmas}

We now turn to the specific example of a hot plasma close to thermal
equilibrium. `Hot' implies that particle masses can be neglected to
leading order $m\ll T$, and that the gauge coupling, as a function of
temperature, is very small $g(T)\ll 1$. To simplify the analysis we
shall perform some approximations, all of which can be understood as a
systematic expansion in $g$ (and $\epsilon$). After ensemble
averaging, the distribution function $f$ is effectively coarse-grained
over a Debye volume. Fluctuations of $\bar f$ within a Debye volume
are parametrically suppressed by powers of $g$. We expand $f$ about
the equilibrium distribution function $\bar f^{\rm eq}$ to leading
order in $g$ as
\beq\label{Ansatz} 
f=\bar f^{\rm eq}+g\bar f^{(1)}+\delta f\ .  
\eeq
Solving (\ref{NA-1}) in the {\it first moment approximation}, that is
using (\ref{Ansatz}) for $\delta f\equiv 0$, has been shown in
ref.~\cite{KLLM} to reproduce the HTL effective theory.

Beyond the HTL level we employ the {\it polarisation approximation}
where $\delta f\neq 0$. It consists in discarding cubic correlator
terms in (\ref{NA-1}) in favour of quadratic ones. Employing
(\ref{polarisation}) means that the effects of collisions are
neglected for the dynamics of the fluctuations. All these
approximations can be systematically improved to higher order.
  
Let us consider colour excitations, described by the colour current density
\beq\label{Jdens}
{\cal J}^\mu_a(x,{\bf v})=
\frac{g\, v^\mu}{\pi^2}\,\int dQ\,dp_0\,d|{\bf p}|\,|{\bf p}|^3\, 
\Theta(p_0)\,\delta(p^2)\, Q_a\, f(x,p,Q)\ ,
\eeq
and $v^\mu\equiv p^\mu/p_0=(1,{\bf v}), {\bf v}^2=1$. The current $J$
is obtained after an angle average over the directions of ${\bf v}$ as
$J(x)=\int\frac{d\Omega_{\bf v}}{4\pi}{\cal J}(x,{\bf v})$. The colour
measure $dQ$ is normalised $\int dQ=1$ and contains the Casimir
constraints of the gauge group, such as $\int dQQ_aQ_b=C_2\delta_{ab}$
with the quadratic group Casimir $C_2=N$ for particles in the 
adjoint representation of $SU(N)$ and $C_2=\frac{1}{2}$ for particles 
in the fundamental \cite{LM2,KLLM}. The approximate dynamical 
equations for the
fluctuations become
\begin{eqnarray}
&&\left(v^\mu \bar D_\mu \,  \delta   {\cal J}^\rho \right)_a \, = 
-m^2_D v^\rho v^\mu 
\left(\bar D_\mu a_0-\bar D_0 a_\mu\right)_a   -g f_{abc}  v^\mu a_\mu ^b
\bar {\cal J}^{c,\rho} 
\ ,\label{vD-dJa} \\
&&\left( \bar D^2 a^\mu-\bar D^\mu(\bar D a)\right)_a+2 g f_{abc} 
\bar F_b^{\mu\nu}a_{c,\nu}=\delta J_a^\mu  \ ,\label{dJa}
\end{eqnarray}
where the quantum Debye mass for $SU(N)$ with $N_F$ quarks and anti-quarks 
\begin{eqnarray} \nonumber
m^2_D&=&-\frac{g^2}{\pi^2}\int^\infty_0dp p^2
       \left( N  \frac{\bar f^{\rm eq}_B}{dp}
             +N_F\frac{\bar f^{\rm eq}_F}{dp} \right)
\\ &=&g^2T^2(2N+N_F)/6
\end{eqnarray} 
obtains from the equilibrium (Bose-Einstein or Fermi-Dirac)
distribution function and the group representation of the particles.
The classical Debye mass follows by using the Maxwell-Boltzmann
distribution, instead. Solving for the fluctuations in the present
approximations yields a closed expression for $\delta {\cal J}$, and
an iterative expansion in powers of the background fields for
$a$ (see ref.~\cite{LM2} for the explicit expressions). 
The seeked-for dynamical equation for the mean
quasi-particle colour excitations is
\begin{eqnarray} \label{NAV-soft-mean}
v^\mu \bar D_\mu \bar {\cal J}^0+ m^2_D v^\mu \bar F_{\mu0}
&=& C_{\rm lin}[\bar {\cal J}^0] +\zeta(x,{\bf v})\ ,\label{NAV-curr}
\end{eqnarray}
where the linearised collision integral $C_{\rm lin}$ can now be
evaluated explicitly, using the expressions for $a$ and $\delta {\cal
  J}$. The Yang-Mills equation remains unchanged at this order. For
the collision integral, one finally obtains to linear order in the
mean current, and to leading logarithmic order (LLO) in $(\ln
1/g)^{-1}\ll 1$
\begin{eqnarray}\nonumber
C_{\rm lin}[\bar {\cal J}_a^0](x,{\bf v})&=&
\left.gf_{abc}\left\langle a^b_\mu(x)\,
\delta {\cal J}^{c,\mu}(x,{\bf v})\right\rangle
\right|_{\bar A=0,\ {\rm linear \, in\,}\bar{J},\ {\rm LLO}}\\ \label{Clin}
&=&
-\gamma\int\frac{d\Omega'}{4\pi}I({\bf v},{\bf v}')\bar{\cal J}_a^0(x,{\bf v}')
\end{eqnarray}
with the kernel $I({\bf v},{\bf v}')=\delta^2 ({\bf v}-{\bf
  v}')-\frac{4}{\pi}({\bf v.v'})^2/\sqrt{1-({\bf v.v'})^2}$. Notice
that the collision integral is local in coordinate space, but
non-local in the angle variables. The rate
$\gamma=\frac{g^2T}{4\pi}\ln 1/g$ is (twice) the hard gluon damping
rate. In obtaining (\ref{Clin}), we introduced an infra-red cut-off
$\sim g m_D$ for the elsewise unscreened magnetic sector. This result
has been first obtained by B\"odeker in ref.~\cite{DB}. 
For alternative derivations, see ref.~\cite{ASY,BI}. In addition,
the source for stochastic noise $\zeta$ in (\ref{NAV-soft-mean}) can be
identified as
\beq\label{zeta}
\zeta_a(x,{\bf v})=
\left. gf_{abc} a^b_\mu(x)\,
\delta {\cal J}^{c,\mu}(x,{\bf v})\right|_{\bar A=0,\,\bar J=0}\ .
\eeq
Making use of the basic correlator
(\ref{BasicGibbs}), its self-correlator follows to LLO as
\beq
\label{noise1}
\langle \zeta^a(x,{\bf v})\zeta^b(y,{\bf v}')\rangle =
2\gamma \,m_D^2 \,T \,I({\bf v},{\bf v}')\,
\delta^{ab}\,\delta (x-y)\ .
\eeq
Notice that the strength of the correlator is determined by the kernel
of the collision integral, e.g.~by the dissipative process. Next, we
show the close relationship to the fluctuation-dissipation theorem
(FDT).

\section{Fluctuation-dissipation theorem}

It is well-known that dissipation in a quasi-stationary plasma (= no
entropy production) is intimately linked to the fluctuations, a link
which is given by the FDT. We shall argue that the FDT holds true for
the above set of equations. To that end, following ref.~\cite{LM3}, we
consider the coarse-grained kinetic equation (\ref{NA-f-cg}) and ask
how the spectral functions of the noise source $\zeta$ and of $f$ have
to be related to $C[f]$ in order to satisfy the FDT.

Within classical transport theory, the FDT is implemented in a
straightforward way \cite{LL}. The pivotal element is the kinetic
entropy $S[f]$. We consider small deviations of $f$ from the
equilibrium, $f=f^{\rm eq}+\Delta f$. The entropy, stationary at
equilibrium, is expanded to quadratic order in $\Delta f$, $S=S_{\rm
  eq}+\Delta S$. Defining the thermodynamical force in the usual way
as $F(z)=-\delta (\Delta S)/\delta (\Delta f(z))$, where $z\equiv
(x,{\bf p}, Q)$, we obtain
\beq\label{Force}
F(x,{\bf p},Q)=\Delta f(x,{\bf p},Q)/\bar f_{\rm eq}
\eeq
for a classical plasma. The thermodynamical force for the quantum 
case follows from (\ref{Force}) using (\ref{replace}). 
Given $C[f](z)$, it can be shown \cite{LM3,LL} that 
\beq\label{noisenoise}
\langle\zeta(z)\zeta(z')\rangle=
-\left( \frac{\delta C(z)}{\delta F(z')}
       +\frac{\delta C(z')}{\delta F(z)}\right)\ 
\eeq
is the required noise-noise self-correlator for $\zeta$ compatible
with the FDT. For the particular example studied above, (\ref{noise1})
follows from inserting (\ref{Clin}) into (\ref{noisenoise}), proving
that B\"odeker's effective theory is compatible with the FDT.
Furthermore, the correlator $\langle\Delta f\Delta f\rangle|_{t=t'}$
can be derived along the same lines and agrees, to leading order, with
$\langle\delta f\delta f\rangle|_{t=t'}$ derived from the Gibbs
ensemble average. This guarantees that the formalism is consistent
with FDT.

\section{Summary}

We have reviewed a semi-classical approach to derive effective
transport equations for QCD based on `integrating-out' fluctuations
about some mean fields. Most interestingly, the approach is
applicable for in- and out-of-equilibrium situations, opening a
door for future applications to out-of-equilibrium plasmas. The
formalism is consistent with the underlying non-Abelian mean field
symmetry, and allows for systematic expansions controlled by a small
plasma parameter which is at the basis for any reliable computation.

For thermal plasmas we have discussed how the collision integral as
well as the necessary noise source follow explicitly from the
microscopic theory to leading logarithmic accuracy, reproducing
B\"odeker's effective theory, and establishing the consistency with
the fluctuation-dissipation theorem.

As a final comment we point out that the effective theory is the same
for a classical or a quantum plasma, differing only in the equilibrium
distribution function, and hence in the corresponding value for the
Debye mass. The sole `quantum' effect which entered the computation
resides in the non-classical statistics of the particles, which is all
that is needed to correctly describe a hot {\it quantum} non-Abelian
plasma close to equilibrium at the present order of accuracy.

\end{document}